\documentstyle[12pt,a4]{article}
\begin{document}
\title{A QUASIPERIODIC GIBBONS--HAWKING METRIC AND SPACETIME FOAM}
\author{Serdar Nergiz and Cihan Sa\c{c}l\i o\~{g}lu\thanks{Permanent addresss:
Physics Department, Bo\~{g}azi\c{c}i University,
80815 Bebek--\.{I}stanbul, Turkey} }
\date{Physics Department, Bo\~{g}azi\c{c}i University \\
80815 Bebek--\.{I}stanbul, Turkey \\  and \\ Physics Department, TUBITAK
\\  Marmara Research Center \\
Research Institute for Basic Sciences \\ 41470 Gebze, Turkey}
\maketitle
\vspace*{1cm}
\begin{abstract}
We present a quasiperiodic self-dual metric of the Gibbons--Hawking type
with one gravitational instanton per spacetime cell. The solution, based on
an adaptation of Weierstrassian $\zeta$ and $\sigma$ functions to three
dimensions, conforms to a definition of spacetime foam given by Hawking.
\end{abstract}
\vspace*{5 cm}
\pagebreak
\baselineskip=30pt

\noindent \bf 1. Introduction \rm

Recently there has been much interest in studying integrability properties
\cite{art1,art2} and underlying infinite dimensional algebras
\cite{art3,art4,art5} of self-dual Einstein's equations. Earlier work on
self-dual metrics, on the other hand, was focused primarily on their
suspected relevance to the quantization of gravity
\cite{art6,art7,art8,art9} and/or their possible effects in baryon/lepton
conservation \cite{art9,art10}. In particular, Hawking \cite{art9} argued
that the dominant contribution to $N(V)dV$, the number of gravitational
fields with compactified spacetime volumes between $V$ and $V+dV$, comes
from metrics containing one gravitational instanton per characteristic
volume, whose size is defined by a normalization constant, presumably
related to Planck mass. According to Hawking, such metrics result in a
foam-like structure of spacetime. The main purpose of this note is to offer
an explicit example of a spacetime foam metric based on an infinite-center
generalization of the Gibbons--Hawking \cite{art11} solution.

Hawking's description of spacetime foam indicates that such a metric will
involve periodic or at least quasiperiodic functions of the coordinates.
As Gibbons--Hawking metrics exhibit similarities to
Jackiw--Nohl--Rebbi--`t~Hooft \cite{art12} Yang--Mills multi-instanton
solutions, periodic versions of the latter can be useful in providing
insights for constructing periodic versions of the former. The first example
of a periodic Yang--Mills instanton solution is due to Rossi \cite{art13},
who considered an infinite string of equal size and equally spaced
Jackiw--Nohl--Rebbi instantons arranged along the Euclidean time axis. In
this case, the periodic time dependence surprisingly turns out to be a gauge
artifact and the solution is seen to be a static BPS monopole \cite{art14},
with $A^{a}_{4}$ playing the role of the Higgs field $\varphi^{a}$. The
mass of the monopole is simply the action or topological charge per unit
time.

G\"{u}rsey and Tze \cite{art15} took this further by writing down a
self-dual Yang--Mills connection with unit instanton per spacetime cell.
In the light of Rossi's result, it is natural to interpret this solution as
one representing BPS monopoles arranged on a three dimensional lattice
\cite{art16}. As the lattice separation becomes smaller, such a
configuration can be viewed as the much sought after monopole condensate
\cite{art17}, or, as the Yang--Mills counterpart of spacetime foam. Such
considerations suggest that the semiclassical model for the true ground
state (rather than the perturbation theory vacuum) in both General
Relativity and Yang--Mills theory consists of a coherent superposition of
the instantons of the theory.

One of the respects in which the two problems are dissimilar, however, is
the fact that gravitational instantons, unlike Yang--Mills ones, have zero
action. Thus their contribution to the path integral is not supressed by the
Boltzmann factor, hinting that the role of instantons in gravity may be
even more fundamental than in Yang--Mills theory.

In mathematical terms, the G\"{u}rsey--Tze solution is based on Fueter's
quaternionic generalizations $\zeta^{F}$ and $\sigma^{F}$ of the
Weierstrassian quasiperiodic functions $\zeta (z)$ and $\sigma (z)$.
However, the solution can be written also in a quaternion-free form
\cite{art16} which allows one to extend Weierstrassian functions to any
dimension; and in particular, to the triply quasiperiodic Gibbons--Hawking
$V(\vec{r})$ used here.

We review the definitions and relevant properties of $\zeta (z)$ and
$\zeta^{F}$ in Section 2. We construct $V(\vec{r})$ and study its
transformation under lattice shifts in Section 3. Topological numbers are
discussed in Section 4. After some concluding observations in Section 5, we
present the magnetic monopole vector potential $\vec{\omega}(\vec{r})$, in
the Appendix.

\pagebreak \noindent \bf 2. Complex and quaternionic quasiperiodic functions:
\rm

The Gibbons--Hawking k--instanton metric has the form
\begin{equation}\label{1}
ds^{2} = \frac{1}{V} (d\tau + \vec{\omega} \cdot d\vec{r} )^{2} +
V d \vec{r} \cdot d\vec{r}
\end{equation}
with
\begin{equation}\label{2}
V = \sum_{i=1}^{k+1} \frac{1}{|\vec{r}-\vec{r}_{i} \, | }~~.
\end{equation}
(Anti) Self-duality is imposed by choosing an $\vec{\omega}$ such that
\begin{equation}\label{3}
\vec{\nabla} \times \vec{\omega}(\vec{r})=\pm \vec{\nabla}V(\vec{r})~~.
\end{equation}

The variable $\tau$ is restricted to $[0, 4\pi]$. Around each
singularity $\vec{r}=\vec{r}_{i}$ of $V(\vec{r})$, one chooses a different
patch and different patches are related by
\begin{equation}\label{4}
\tau_{n+1} = \tau_{n} + 2\phi_{n}~~,
\end{equation}
where $\phi_{n}$ is the azimuthal angle with respect to the ``$z$-axis''
along $(\vec{r}_{n+1} - \vec{r}_{n})$.

Our aim is now to construct a $V(\vec{r})$ for which the $\vec{r}_{i}$ are
points $ \{ \vec{q}\, \}$ belonging to a three dimensional lattice. This
means the $\vec{q}$ have the form
\begin{equation}\label{5}
\vec{q} =
n_{1} \vec{q}^{\, (1)} + n_{2} \vec{q}^{\, (2)} + n_{3} \vec{q}^{\, (3)}~~,
\end{equation}
where the $\vec{q}^{\, (a)}$ ($a= 1,2,3$) are basis vectors of the
lattice and $n_{a} \in \bf Z \mit$.

One might be tempted to write simply
\begin{equation}\label{6}
V = \sum_{n_{1}} \sum_{n_{2}} \sum_{n_{3}}
\frac{1}{|\vec{r}-\vec{q} \, |}~~,
\end{equation}
but this expression is not convergent: An integral version of (\ref{6})
exhibits a quadratic divergence of the form
\begin{equation}\label{7}
\int_{|\vec{q} \, |_{min}}^{\infty} d|\vec{q} \, | \, |\vec{q} \, |^{2} /
|\vec{q} \, |~~.
\end{equation}

A similar problem in the definition of Weierstrassian elliptic functions is
solved by subtraction terms. For example, in
\begin{equation}\label{8}
\wp (z) = \frac{1}{z^{2}} + \sum_{\vec{n} \neq 0} \{ \frac{1}{(z-\omega)^{2}}
-\frac{1}{\omega^{2}} \} ~~,
\end{equation}
where $\omega = n_{1} \omega_{1} + n_{2} \omega_{2} $, in the first term
one encounters a logarithmic divergence basically of the form
\begin{equation}\label{9}
\int_{\omega_{min}}^{\infty} d|\omega | \, |\omega | / |\omega |^{2} ~~.
\end{equation}
Note that since the series (\ref{8}) is not absolutely convergent, the
parentheses are essential to its definition. Similarly, the function
\begin{equation}\label{10}
\zeta (z) = - \int \wp (z) dz = \frac{1}{z} + \sum_{\vec{n} \neq 0}
\{ \frac{1}{(z-\omega)} + \frac{1}{\omega} + \frac{z}{\omega^{2}} \}
\end{equation}
necessitates two subtraction terms. Note again the importance of the
brackets without which the $1/ \omega $ term would sum to zero by itself,
rendering the sum meaningless. The manifestly convergent form of
\cite{art10} is
\begin{equation}\label{11}
\zeta (z) = \frac{1}{z} + \sum_{\vec{n} \neq 0} \frac{z^{2}}{\omega^{2}
(z-\omega)}~~.
\end{equation}
Finally, the $\sigma$ function, whose logarithmic derivative equals
$\zeta (z)$, is defined via
\begin{equation}\label{12}
\phi \equiv \ln \sigma (z) =\ln z + \sum_{\vec{n} \neq 0} \{ \ln
(z-\omega) -  \ln (-\omega) + \frac{z}{\omega} + \frac{z^2}{2 \omega^2}\}~~.
\end{equation}

We will base our generalization on the following properties of $\phi (z)$ :
$$(i)~~~\partial_{z} \partial_{\overline{z}} \phi \propto \sum{} \sum{}
\delta (x-\omega_{x}) \delta (y-\omega_{y})~~, \eqno(13.a)$$
$$(ii)~~~\partial_{z} \partial_{\overline{z}} \{ -\ln (-\omega)+\frac{z}
{\omega}+\frac{z^2}{2 \omega^2}\} = 0~~, \eqno(13.b)$$
$(iii)$ three subtraction terms (up to $z^2/ \omega^2$) are needed
such that for $|z| \ll |w|_{min}$,
$$\phi (z) \cong \ln z + \sum{} \sum{} O(\frac{z^3}{\omega^3}) +
O(\frac{z^4}{\omega^4}) + \cdots ~~.  \eqno(13.c)$$
The last property is dictated by convergence requirements and simple
dimensional analysis: the terms in the sum cannot involve powers of $\omega$
higher than $\omega^{-3}$; the dimensionlessness of $\phi (z)$ then implies
the form (13.c).

Fueter has constructed analogues of (\ref{11}) and (\ref{12}) by using a
quaternionic variable $\bf x\mit \equiv Ix_{0}-i\vec{\sigma} \cdot \vec{x}$
instead of $z$. However, a suitable adaptation of (13) allows the definition
of quasiperiodic functions over any $\bf R^{n}\mit$, suggesting quaternionic
techniques are not essential to the construction in $\bf R^{4}\mit$. Indeed,
using (13) with
$ q_{\mu} = n_{0} q_{\mu}^{(0)}+n_{1} q_{\mu}^{(1)}+n_{2} q_{\mu}^{(2)}+
n_{3} q_{\mu}^{(3)}$,
one finds \cite{art10}, \setcounter{equation}{13}
\begin{equation}\label{14}
\rho (x) = \frac{1}{x^{2}} +\sum_{\{ q \} \neq 0} \{
\frac{1}{(x-q)^{2}} -\frac{1}{q^{2}} -\frac{2 q \cdot x}{q^{4}} -
\frac{1}{q^{6}} (4(q \cdot x)^{2} - q^{2} x^{2}) \}
\end{equation}
for the four-dimensional counterpart of $\phi (z)$. The analogue of
(\ref{11}) is the vector field defined by
\begin{equation}\label{15}
\zeta_{\mu}^{F} (x) = \partial_{\mu} \rho (x)~~.
\end{equation}
In Fueter's quaternionic notation this is converted to the quaternion-valued
function
\begin{equation}\label{16}
\bf \zeta^{\mit F} \mit (x) = I\partial_{0}\rho + i\vec{\sigma} \cdot
\vec{\nabla} \rho
\end{equation}
which restores the formal similarity to (\ref{11}).

\noindent \bf 3. The triply quasiperiodic Gibbons--Hawking potential: \rm

It is now a straightforward matter to apply (13.a,b,c) in $\bf R^{3}\mit$ to
obtain
\begin{equation}\label{17}
V(\vec{r}) = \frac{1}{r} + \sum_{} \sum_{ \{ \vec{q} \} \neq 0 } \sum_{} \{
\frac{1}{|\vec{r}-\vec{q} \, | } - \frac{1}{|\vec{q} \, | } [1+
\frac{\vec{q} \cdot \vec{r}}{q^{2}} + \frac{1}{2q^{4}}
(3(\vec{q} \cdot \vec{r})^{2} - q^{2} r^{2}) ] \}~~.
\end{equation}
Note again that (\ref{14}) and (\ref{17}) are only meaningful if the outer
parentheses are respected. The last two terms in both (\ref{14}) and
(\ref{17}) cannot be separately summed to zero anymore than the $1/ \omega $
in (\ref{10}) can.

The analogue of $\zeta (z)$ is now the vector field
$\vec{\nabla} V(\vec{r})$. It is well known that under lattice shifts,
$\zeta$ obeys the quasiperiodic transformation law
\begin{equation}\label{18}
\zeta (z + \omega_{1,2}) = \zeta (z) + \eta_{1,2}~~,
\end{equation}
where $\eta_{1,2}$ are constant complex numbers obeying Legendre's relation
\begin{equation}\label{19}
\eta_{1} \omega_{2} - \eta_{2} \omega_{1} = 2 \pi i~~.
\end{equation}
Integrating (\ref{18}) and using the oddness of $\sigma (z)$ one has
\begin{equation}\label{20}
\sigma (z + \omega_{1,2}) = -\sigma (z)  \exp [\eta_{1,2} (z + \frac{1}{2}
\omega_{1,2})]~~.
\end{equation}
To derive $\bf R^{3}$ generalizations of (\ref{18})-(\ref{20}) we note
$\nabla^{2} V$ is a perfectly triply periodic arrangement of $\delta$-
functions; thus we may integrate it once to obtain
\begin{equation}\label{21}
\vec{\nabla} V(\vec{r} + \vec{q}^{\, (a)}) = \vec{\nabla} V(\vec{r}) +
\vec{\eta}^{\, (a)}~~,~~(a=1, 2, 3)~~,
\end{equation}
where the $\vec{\eta}^{\, (a)}$ are constant vectors. Integrating once more
and using the fact that $V(-\vec{q}^{\, (a)}/2) = V(\vec{q}^{\, (a)}/2)$,
which incidentally means
$\eta_{i}^{\, (a)} =2(\partial_{i} V)_{\vec{r}=\vec{q}^{\, (a)}/2}$, we find
\begin{equation}\label{22}
V(\vec{r} + \vec{q}^{\, (a)}) = V(\vec{r}) +
\vec{\eta}^{\, (a)} \cdot (\vec{r} + \frac{\vec{q}^{\, (a)}}{2})~~
(\rm no~sum~over~\mit a)~~.
\end{equation}
This is clearly the counterpart of (\ref{20}). Integrating $\vec{\nabla} V$
over the surface ($\partial \rm cell$) of a period cell yields
\begin{equation}\label{23}
\sum_{a=1}^{3} \sum_{b=1}^{3} \sum_{c=1}^{3} \frac{1}{2} \epsilon^{abc}
\vec{\eta}^{\, (a)} \cdot (\vec{q}^{\, (b)} \times \vec{q}^{\, (c)})
= -4 \pi~~.
\end{equation}
This replaces Legendre's relation (\ref{19}) in three dimensions. We may
also iterate (\ref{22}) to find
\begin{equation}\label{24}
V(\vec{r} + m \vec{q}^{\, (a)}) = V(\vec{r}) +
\vec{\eta}^{\, (a)} \cdot (m \vec{r} + \frac{m^2}{2} \vec{q}^{\, (a)})~~.
\end{equation}

\noindent \bf 4. Topological numbers: \rm

As we are dealing with self-dual curvature two-forms $R^{a}_{~b}$, the
Euler class
\begin{equation}\label{25}
e = \frac{1}{32\pi^2} \epsilon_{abcd} R^{a}_{~k} \wedge R^{c}_{~l}
\delta^{bk} \delta^{ld}
\end{equation}
and the Pontrjagin class
\begin{equation}\label{26}
p_{1} = -\frac{1}{8\pi^2}  R^{a}_{~b} \wedge R^{b}_{~a}
\end{equation}
are simply proportional to each other. The topological numbers related to
(\ref{25}) and (\ref{26}) are the Euler characteristic $\chi$ and the
signature $\tau$, respectively. For a $(k+1)$-center Gibbons--Hawking metric
one has \cite{art19,art20}
\begin{equation}\label{27}
\tau_{k} = \frac{1}{3} \int p_{1} - \frac{1}{k+1} \sum_{n=1}^{k} \cot^{2}
(\frac{n\pi}{k+1}) = -k
\end{equation}
and
\begin{equation}\label{28}
\chi = k+1~~.
\end{equation}
Remarkably, the analogy between Jackiw--Nohl--Rebbi--`t~Hooft and
Gibbons--Hawking
instantons carries over to the expressions for the topological
charges. In the Yang--Mills case, the connection
\begin{equation}\label{29}
A_{\mu} = i \overline{\sigma}_{\mu \nu} \partial_{\nu} \ln \rho
\end{equation}
with
\begin{equation}\label{30}
\rho = \sum_{i=1}^{n+1} \frac{\lambda_{i}^{2}}{(x-x_{i})^{2}}
\end{equation}
gives rise to a self-dual field strength, which results in the topological
charge density
\begin{equation}\label{31}
-\frac{d^{4}x}{16\pi^{2}}Tr(\tilde{F}_{\mu \nu} F_{\mu \nu}) =
-\frac{d^{4}x}{16\pi^{2}} \Box \Box \ln \rho~~.
\end{equation}
The integral of this expression gives n when converted into a surface
integral over a large $S^{3}$ containing all the singularities plus
infinitesimal $S^{3}$'s around each singularity. In the Gibbons--Hawking
parametrization, a lengthy but straighforward calculation yields
\begin{equation}\label{32}
p_{1} = -\frac{1}{16\pi^{2}}  d\tau \wedge dx^{1} \wedge dx^{2}
\wedge dx^{3} \, \triangle \triangle \frac{1}{V}~~.
\end{equation}
Just as in the Yang--Mills computation based on (\ref{31}), one can convert
the volume integral of (\ref{32}) to the surface integral
\begin{equation}\label{33}
\frac{1}{3} \int p_{1}  = -\frac{1}{48\pi^{2}} \int d\tau \int d\vec{\sigma}
\cdot \vec{\nabla}(\frac{2\vec{\nabla}V\cdot \vec{\nabla}V}{V^{3}})=
-\frac{1}{6\pi^{2}} \int d\vec{\sigma}
\cdot \vec{\nabla}(\frac{\vec{\nabla}V\cdot \vec{\nabla}V}{V^{3}})~~,
\end{equation}
where the surface consists of a large $S^{2}$ containing all the
singularities of $V$ plus $k+1$ infinitesimal $S^{2}$'s around each
singularity. The result is
\begin{eqnarray}\label{34}
\tau_{k}
& = &
\frac{2}{3}(\frac{1}{k+1}-(k+1))- \frac{1}{k+1} \sum_{n=1}^{k}
\cot^{2}(\frac{n\pi}{k+1})
\nonumber \\
& = &
\frac{2}{3}(\frac{1}{k+1}-(k+1))-\frac{k(k-1)}{3(k+1)} = -k
\end{eqnarray}
as expected. For the solution corresponding to (\ref{17}) with a singularity
per unit cell, $\tau(\rm per~cell\mit) = - 1$. Although we have been working
with a lattice
in $\bf R^{3}$, the fourth variable is already periodic with a period $4\pi$;
hence this is to be regarded as topological number per \underline{spacetime}
cell. Thus we have a close analogue of the self-dual Yang--Mills solution
due to G\"{u}rsey and Tze \cite{art15}.

\pagebreak \noindent \bf 5. Concluding remarks: \rm

The quasiperiodic functional form given in (\ref{17}) defines a different
metric for each choice of three dimensional lattice. Furthermore, just as
one can superpose k terms of the form $| \vec{r} - \vec{r}_{i} \, |^{-1}$,
one can superpose $V$'s for different lattices, or for isometric lattices
with periods that are multiples of some irreducible basis vectors. Solutions
such as (\ref{17}) can then be used as a building blocks for more complex
types of space-time foam. On the other hand, there must exist another class
of manifolds with similar periodic properties, but which cannot be obtained
from (\ref{17}). These correspond to asymptotically locally Euclidean
self-dual manifolds whose boundaries are other spherical forms of $S^{3}$
related to the dihedral groups $D_{k}$ of order $k$ and to certain discrete
groups. Since these metrics are not known, the treatment in this paper
cannot at present be extended to the corresponding instantons.We will
nevertheless venture some speculations concerning the lattices that are
likely to be encountered.

It is natural to expect that lattices that correspond to the tightest
packing of spheres might play a special role. For example, in the
Copenhagen model \cite{art21} for the Yang--Mills vacuum one first considers
two dimensional lattices of chromomagnetic vortex tubes, which yield a
vacuum energy below that of the perturbation theory vacuum when one loop
corrections are taken into account. The energy is then lowered further when
the tightest packing corresponding to the hexagonal root lattice of
$SU(3)$ is chosen.

In our four dimensional problem the tightest packing lattice is the root
lattice of $SO(8)$. However, the Dynkin diagram of
$SO(8)$ makes it clear that this is not a possible choice for our
Gibbons--Hawking
class of metrics based on the cyclic groups $A_{k}$. The reason for
this is obvious: the lattice vector in the $\tau$ direction is orthogonal
to all the $\vec{q}^{\, (a)}$ while none of the $SO(8)$ simple roots has
this property. The tightest packing available for the $A_{k}$ class
considered here obtains when the $\vec{q}^{\, (a)}$ are taken as the simple
roots of $SU(4)$. We conjecture that the $SO(8)$ lattice may be relevant for
metrics based on the $D_{k}$ family. If true, this would imply that the
$D_{k}$ metrics (unlike $A_{k}$ ones) cannot be parametrized in terms of
functions independent of $\tau$ such as $V(\vec{r})$ and
$\vec{\omega}(\vec{r})$. Pursuing the analogy between
Jackiw--Nohl--Rebbi--`t~Hooft and Gibbons--Hawking instantons, it is
tempting to regard the $D_{k}$ metrics as analogues of the ADHM \cite{art22}
instanton solution for which the group space orientations of the instantons
are in general not parallel.

\noindent \bf Appendix: \rm

We work in a gauge where the monopole centered at the point
$\vec{q} = n_{1} \vec{q}^{\, (1)} + n_{2} \vec{q}^{\, (2)} +
n_{3} \vec{q}^{\, (3)}$ has a string originating from $\vec{q}$ and lying
parallel to the negative $z$-axis. In order to avoid singularities, the
$z$-axis must not be aligned with any of the $\vec{q}$; the reason for this
restriction will be apparent from the expression below. The construction of
the vector potential $\vec{\omega}$ then proceeds along the lines leading to
(\ref{17}). Thus one starts with
$$\hat{z} \times \{ \frac{\vec{r}}{r} \, \frac{1}{z+r} +
\sum{} \sum_{ \{ \vec{q} \neq 0 \} } \sum{} \frac{\vec{r}-\vec{q}}
{| \vec{r} - \vec{q} \, | } \, \frac{1}
{(z-q_{z})+| \vec{r} - \vec{q} \, | } \}~~, \eqno(A.1)$$
which represents monopole vector potentials centered at the origin and at
the points $\{ \vec{q} \}$. One then Taylor expands the terms in the sum for
small $(x, y, z)$ up to and including terms of quadratic order. The terms
so obtained are then subtracted from (A.1), yielding the expression
$$\vec{\omega} = \hat{k} \times \{ \frac{\vec{r}}{r} \, \frac{1}{z+r} +
\sum{} \sum_{ \{ \vec{q} \neq 0 \} } \sum{} [ \frac{\vec{r}-\vec{q}}
{| \vec{r} - \vec{q} \, | } \, \frac{1}
{(z-q_{z})+| \vec{r} - \vec{q} \, | } - \vec{\Omega}_{q} ] \}~~, \eqno(A.2)$$
where
\begin{eqnarray*}
\vec{\Omega}_{q}
& = &
\frac{1}{q(q-q_{z})} \{
- \vec{q} + (
\frac{1}{q^{2}} [\vec{r} q^{2}- \vec{q} (\vec{q} \cdot \vec{r})] -
\frac{\vec{q}[(\vec{q} \cdot \vec{r}) - zq]}{q(q-q_{z})}
)
\nonumber \\
&  & \mbox{}
+ \frac{1}{2} [
\vec{q} \, \frac{r^{2}}{q^{2}}+2\vec{r}\, \frac{(\vec{q}\cdot \vec{r})}{q^{2}}-
3\vec{q} \, \frac{(\vec{q}\cdot \vec{r})^{2}}{q^{4}} +
2[\vec{r} q^{2}- \vec{q} (\vec{q} \cdot \vec{r})]
\frac{[(\vec{q} \cdot \vec{r}) - zq]}{q^{3}(q-q_{z})}
\nonumber \\
&  & \mbox{}
- \vec{q} (
- \frac{[\hat{k}\cdot (\vec{r} \times \vec{q})]^{2}}{q^{3}(q-q_{z})} +
\frac{2(\vec{q}\cdot \vec{r})^{2}}{q^{2}(q-q_{z})^{2}} -
\frac{2z(\vec{q}\cdot \vec{r})}{q^{3}(q-q_{z})^{2}}
(q_{z}^{2}+2q^{2}-qq_{z})
\nonumber \\
&  & \mbox{}
- \frac{r^{2} q_{z}^{2}}{q^{3}(q-q_{z})} +
\frac{z^{2} (q-q_{z})}{q^{3}} +
\frac{z^{2} q_{z}}{q^{3}(q-q_{z})^{2}}
(q_{z}^{2}+4q^{2}-3qq_{z})
) ] \}~~.~~~~~~~~~~~~~~~~~~~~~(A.3)
\end{eqnarray*}

It is not difficult to verify that this $\vec{\omega}$ satisfies
$$\vec{\nabla} \times \vec{\omega} = \vec{\nabla}V \eqno(A.4)$$
for the $V$ given in (\ref{17}).

{\bf Acknowledgements}

We thank Alikram Aliyev, Yavuz Nutku, Rahmi G\"{u}ven and our other Theory
Group colleagues
in Gebze and Bo\~{g}azi\c{c}i for useful discussions and encouragement.
C. S. gratefully acknowledges the hospitality and support extended to him at
the Marmara Research Center. This work was partially supported by the
Turkish Scientific and Technical Research Council, TBAG \c{C}G-1.


\begin{thebibliography}{99}

\bibitem{art1} I.A.B. Strachan, "The symmetry structure of the anti-self-dual
Einstein hierarchy", University of Newcastle preprint, 1995.
\bibitem{art2} J.D. Finley, J.F. Pleba\~{n}ski, M. Przanowski,
H. Garc\'{\i}a-Compe\'{a}n,
Phys. Lett. {\bf A 181} (1993) 435.
\bibitem{art3} V. Husain,
Phys. Rev. Lett. {\bf 72} (1994) 800.
\bibitem{art4} H. Garc\'{\i}a-Compe\'{a}n, L.E. Morales and J.F. Pleba\~nski,
preprint CINVESTAV-FIS GFMR 10/94.
\bibitem{art5} J.D.E. Grant,
Phys. Rev. {\bf D 48} (1993) 2606.
\bibitem{art6} A. Ashtekar, T. Jacobson and L. Smolin,
Comm. Math. Phys. {\bf 115} (1981) 631
\bibitem{art7} A. Ashtekar,
Phys. Rev. {\bf D 36} (1987) 1587.
\bibitem{art8} R. Penrose,
Gen. Relativ. Gravit. {\bf 7} (1976) 31.
\bibitem{art9} S. W. Hawking in General Relativity, An Einstein Centenary
Survey, ed. by S. W. Hawking and W. Israel (Cambridge Univ. Press,
Cambridge, England, 1979)
\bibitem{art10} T. Eguchi, and P.G. O. Freund,
Phys. Rev. Lett. {\bf 37} (1976) 1251.
\bibitem{art11} G. W. Gibbons and S. W. Hawking,
Phys. Lett. {\bf 78B} (1978) 430.
\bibitem{art12}R. Jackiw, C. Nohl and C. Rebbi,
In Particles and Fields, ed. by D. H. Boal and A. N. Kamal (Plenum, New York,
1977);
G. `t Hooft, Phys. Rev. {\bf D 14} (1976) 3432.
\bibitem{art13} R. Rossi,
Nucl. Phys. {\bf B 149} (1979) 170.
\bibitem{art14} M. K. Prasad, C. Sommerfield,
Phys. Rev. Lett. {\bf 35} (1975) 760.
\bibitem{art15} F. G\"{u}rsey and H. C. Tze,
Ann. of Phys. {\bf 128} (1980) 29.
\bibitem{art16} S. Nergiz and C. Sa\c{c}l\i o\~{g}lu,
MRC preprint, hep-th/950 3039.
\bibitem{art17}  N. Seiberg and E. Witten,
Nucl. Phys. {\bf B 426} (1994) 19.
\bibitem{art18} R. Fueter,
``Function Theory of a Hypercomplex Variable'', Lectures at the University
of Zurich (1948/49); Congress Int. Math. Oslo {\bf 1} (1936) 75.
\bibitem{art19} G. W. Gibbons and S. W. Hawking,
Comm. Math. Phys. {\bf 66} (1979) 291.
\bibitem{art20} T. Eguchi, P. Gilkey and A. Hanson,
Phys. Rep. {\bf 66} (1980) 213.
\bibitem{art21} H. B. Nielsen in Particle Physics 1980, ed. by I. Andric,
I. Dadic and N. Zovko (North Holland 1981) and references therein; \\
J. Ambjorn and P. Olesen,
Nucl. Phys. {\bf B 170} (1980) 60.
\bibitem{art22} M. F. Atiyah, N. J. Hitchin, V. G. Drinfeld and Yu. I. Manin,
Phys. Lett. {\bf A 65} (1978) 185.

\end{thebibliography}
\end{document}